# An Open and Collaborative Database of Properties of Materials for High-Temperature Superconducting-Based Devices

Pablo Cayado, João Rosas, João Murta-Pina, *Senior, IEEE*, and Harold S. Ruiz

*Abstract*— The successful integration of high-temperature superconductors (HTS) into modern technologies requires consistent, accessible, and comprehensive material data, a need that is currently unmet due to the fragmented and incomplete nature of existing resources. This paper introduces a new collaborative, open-access database specifically designed to address this gap by providing standardized data on HTS materials and crucial auxiliary components for HTS applications. The database encompasses extensive data on structural, cryogenic, electrical, magnetic, and superconducting materials, supporting diverse requirements from HTS modelling to magnet design. Developed through collaborative efforts and organized using an ontology-driven data model, this platform is dynamically adaptable, ensuring that it can grow as new materials and data emerge. Key features include user-driven contributions, peer-reviewed data validation, and advanced filtering capabilities for efficient data retrieval. This innovative database, to the knowledge of the authors, being the largest publicly available for material properties of HTS technologies is positioned as a valuable tool for the HTS community, promoting more efficient research and development processes, accelerating the practical application of HTS, and fostering a collaborative approach to knowledge sharing within the field. The database is available at https://sc.hi-scale.grisenergia.pt/app.

## I. INTRODUCTION

THE successful integration of high-temperature superconductors (HTS) into technological applications—from energy-efficient power systems to advanced medical devices and beyond—depends heavily on the accessibility of accurate, comprehensive, and standardized material data. While scientific literature contains extensive HTS data, this information is often fragmented across publications, formatted inconsistently, or incomplete for engineering purposes. Engineers and researchers, especially those involved in HTS modelling, magnet design, and application development, require easy access to standardized data to optimize these materials effectively. However, the lack of a centralized and user-friendly source for such data has created a significant bottleneck in HTS advancement.

Historically, several databases have been developed to meet these data needs[1–7], each with unique focuses, such as specific materials or particular properties. While these resources represent important steps, they often suffer from limitations, including restricted scope, non-intuitive interfaces, and a lack of comprehensive HTS coverage. Additionally, the proprietary nature of some databases or their emphasis on isolated data sets further compounds accessibility issues. Consequently, designers, researchers, and engineers still face challenges in quickly locating and comparing the specific data they need for HTS applications, particularly as new materials and property measurements emerge.

To bridge this gap, we introduce a novel, open-access database aimed at consolidating and standardizing HTS material properties in an easily navigable, online format. Our database serves as a dynamic repository that extends beyond HTS materials to include auxiliary materials essential for integration in technological applications. By incorporating an intuitive interface and robust search capabilities, this platform enables efficient retrieval of critical data, supporting researchers and engineers in both academic and industrial settings.

A distinguishing feature of this database is its collaborative framework, which allows users not only to access information but also to contribute new data. This approach ensures that the database will continue to evolve, with entries spanning a diverse range of HTS materials, including structural, cryogenic, electrical, magnetic, and superconducting properties. Additionally, each data entry is peer-reviewed or verified,

Manuscrip received November XX, 2024; revised Month DD,YYY; accepted Month DD, YYYY. Date of publication Month DD, YYYY; date of current version Month DD, YYYY. This work was supported by the European Cooperation in Science and Technology, COST Action CA19108 (Hi-SCALE, https://www.cost.eu/actions/CA19108/, https://hi-scale.eu/). We are thankful to all members of this action for their contributions and, specially, Dr. Krastyo Buchkov for his continuous support during the data collection process. *(Corresponding authors: Pablo Cayado, João Murta-Pina).*

P. Cayado is with the Department of Quantum Matter Physics (DQMP) of the University of Geneva, Quai Ernest-Ansermet 24, 1211, Geneva, Switzerland; currently at Physics Department of University of Oviedo, Calvo Sotelo s/n, 33007, Oviedo (e-mail: pablo.cayado@unige.ch).

J. Rosas and J. Murta-Pina are with the NOVA School of Science and Technology, Department of Electrical and Computers Engineering, 2829-516 Caparica, Portugal, and with the Centre of Technology and Systems, UNINOVA, and also with the Intelligent Systems Associate Laboratory (LASI), 2829-516, Caparica, Portugal (e-mail: jrosas@uninova.pt; jmmp@fct.unl.pt).

H.S. Ruiz is with the Green Energy and Transport Research Group at the School of Engineering and Space Park Leicester, University of Leicester, Leicester, LE1 7RH, United Kingdom. (e-mail: dr.harold.ruiz@le.ac.uk).

Color versions of one or more of the figures in this article are available online at http://ieeexplore.ieee.org

ensuring that only high-quality and reliable information is disseminated. Such a community-driven platform will provide a reliable foundation for ongoing research and development efforts, addressing the current limitations in HTS data availability and promoting the broader adoption of these materials.

In the sections that follow, we outline the methodologies used for data collection, organization, and verification. We also describe the structure and function of the web application, the data model, and the protocols established to maintain data quality and encourage community engagement. Through this work, we aim to support the superconductivity community, advancing the practical application of HTS technologies by addressing a critical need for accessible and standardized material data.

Finally, the database is available at https://sc.hi-scale.grisenergia.pt/app.

## II. Data Collection and Web Application

This section describes the approach used to collect bibliographic references, structure them into a coherent data model for HTS-based applications and develop a user-friendly online platform to facilitate access and collaborative contributions. The goal is to provide a dynamic and evolving resource that meets the specific needs of the applied superconductivity community.

### A. Data Collection Framework

The data collection phase was a collaborative effort conducted under the framework of the COST Action CA19108 – High-Temperature SuperConductivity for AcceLerating the Energy Transition (Hi-SCALE). Researchers from several institutions were invited to contribute material property data to establish a foundational dataset for the database. The primary data sources were peer-reviewed scientific publications, complemented by validated information from reputable databases and websites of manufacturers. Each source underwent careful validation to ensure the authenticity and relevance of the information provided, thus maintaining high data reliability.

The selection criteria for references included in the database were designed to ensure consistency and relevance to HTS applications. Publications needed to focus on HTS or report specific, pre-defined properties essential for these technologies. To uphold a high standard of data integrity, all included sources were peer-reviewed, and entries were tagged with Digital Object Identifiers (DOIs) wherever possible to facilitate traceability. This rigorous process enhances the value of the database as a reliable research tool, providing users with traceable data that has undergone scientific scrutiny. However, the authors of the database are not liable for exceptional cases where peer-reviewed data is later retracted, corrected, or disputed by newer publications of equivalent standing.

The data compiled is organized into five primary categories, reflecting the functional roles of materials within HTS applications: structural, cryogenic, electrical, magnetic, and superconducting materials. Each category is further subdivided based on specific material types and properties to ensure comprehensive coverage of the diverse factors critical to HTS integration. For instance, within the structural materials category, data entries include fiber-reinforced composites, metallic alloys, ceramics, and polymers. In the superconductors category, data encompasses various forms such as single crystals, thin films, bulk materials, granular wires, and coated conductors.

Each data entry in the database follows a standardized structure to facilitate uniformity and ease of comparison, consisting of:

- **Publication details**: DOI, year, and title of the publication.
- **Material details**: type of material, specific material, and additional notes.
- **Property data**: measured properties, conditions under which measurement were taken (e.g., temperature and magnetic field), and additional comments to provide context.
- **Supplementary information**: availability of raw data on the source and any relevant additional comments.

This standardized structure ensures that users can efficiently access, compare, and analyze data across different material categories.

It is important to remark that information from existing databases such as ref[6] and ref[7], have been thoroughly analysed, allowing us to extract and filter publicly available information that meet our criteria for inclusion and classification of data in Section III. In this sense, our presented database presents more comprehensive and systematic approach to the characterization of structural, cryogenic, electrical, magnetic, and superconducting materials with their relevant applications for the industry and computational modelling communities.

### B. Database Structure and Model Design

The database is organized to efficiently manage diverse material properties while ensuring ease of access for users. Each material category—including superconducting, cryogenic, electrical, magnetic, and structural materials—is represented as a distinct section in the database, with attributes specifically selected to the type of material and its relevant properties. This modular structure allows the database to be easily expanded as new materials and properties are identified through ongoing research.

To handle the complexity of organizing and updating diverse data entries, the database schema is generated using an ontology-driven approach[8]. The ontology defines the relationships between different material types and their associated properties and is modeled using the JSON (JavaScript Object Notation) format. JSON is a widely used data format that is both human-readable and machine-readable, making it ideal for handling structured data in web applications[9]. By utilizing JSON to define the data model, the database can be dynamically updated to accommodate new material categories or properties without the need for extensive manual intervention.

The implemented flexible and scalable design ensures that the database can grow alongside advancements in HTS research. The ability to incorporate new materials, properties, and measurement data quickly and consistently makes the platform a valuable resource for both academic and industrial applications. Additionally, the structured format of the database facilitates efficient data retrieval and comparison, supporting the community in their efforts to develop and optimize HTS-based technologies.

*C. Collaborative Data Contribution and Verification*

One of the key features of the database is its collaborative framework, which enables members of the superconductivity community to contribute new bibliographic references and update existing entries. This approach leverages the collective expertise of researchers, engineers, and practitioners working with HTS to continuously expand the repository of material properties relevant to HTS-based devices.

All new entries undergo a verification process to maintain the reliability and accuracy of the database. Contributions are reviewed by authorized users with expertise in the field before being made available to the broader community. This validation step ensures that the database remains a trustworthy resource, minimizing the risks of inaccuracies or redundant entries.

The contribution process has been designed to be user-friendly and consistent. Contributors are guided through predefined templates that standardize data entry and ensure uniformity across all records. For example, contributors are prompted to categorize materials accurately and provide key information such as material type, measured properties, and measurement conditions. To maintain data coherence, some fields have dependency relationships—for instance, certain properties are only applicable to specific material types, and the interface adjusts dynamically based on the selections made.

These built-in validation tools and structured templates help to minimize errors during data entry and ensure that the information added to the database is coherent and easily searchable. By fostering community involvement and providing clear guidelines for contributions, the database remains a dynamic and evolving resource that reflects the latest advancements in HTS research and technology development.

Additionally, future updates to the database are planned to be supported by an AI-powered platform currently under development, which will automate the extraction and validation of material data from peer-reviewed literature, further enhancing the scalability and sustainability of the repository.

*D. Web Application Development*

The database is accessible through a web application designed to provide a user-friendly interface that facilitates efficient data retrieval and contribution. The application supports advanced search and filtering options, allowing users to quickly locate relevant material data by filtering based on material type, properties, publication details, and measurement conditions.

The front end of the web application offers a responsive and interactive user experience, while the back end manages data processing and ensures secure interactions with the database. The architecture of the platform was chosen to enhance functionality and maintainability, allowing the system to accommodate future updates and additional features. Technologies as ReactJS, Apache, and PHP, among others, have been used, but these details are outside the scope of this technical note.

The web interface organizes material categories into distinct tabs, making it easy for users to navigate the extensive dataset. Each category—Structural Materials, Cryogenic Materials, Electrical and Magnetic Materials, and Superconductors, see Fig. 1—is further subdivided into specific materials and their associated properties. For instance:

- **Structural Materials**: includes fiber-reinforced composites, metallic alloys, ceramics, and polymers, with properties such as impact strength, tensile fracture strain, and thermal conductivity.
- **Cryogenic Materials**: covers fluids and insulators and heat transfer materials, with properties like latent heat of vaporization at boiling point and specific heat, for fluids, and coefficient of linear thermal expansion and thermal conductivity, for insulators and heat transfer materials.
- **Electrical Materials:** these are divided into conductors and dielectrics, with properties such as electrical conductivity and resistivity for the former, and breakdown voltage and permittivity, for the latter.
- **Magnetic Materials**: includes both permanent magnets and soft magnetic materials. The first include properties as the coercive field strength or magnetic remanence, and the second properties as the grain orientation of relative permeability.
- **Superconductors**: details a range of HTS materials with properties such as critical temperature, critical current density, magnetic susceptibility, and thermal diffusivity.

To enhance usability, the web application provides comprehensive search tools that allow users to refine their queries and locate specific data efficiently. Advanced filtering options help users quickly narrow down search results based on key parameters; an essential feature given the extensive range of data included in the database. Additionally, users can save frequently used search configurations, making it easier to access specific datasets in future sessions.

The application also includes mechanisms for user feedback, enabling community members to suggest improvements to the database structure, data model, or web application features. To further support users, tutorials and a frequently asked questions (FAQ) section are available to provide guidance and ensure a smooth experience for both new and experienced researchers. The goal is to create a platform that not only provides reliable and comprehensive data but also encourages continuous engagement and contributions from the superconductivity

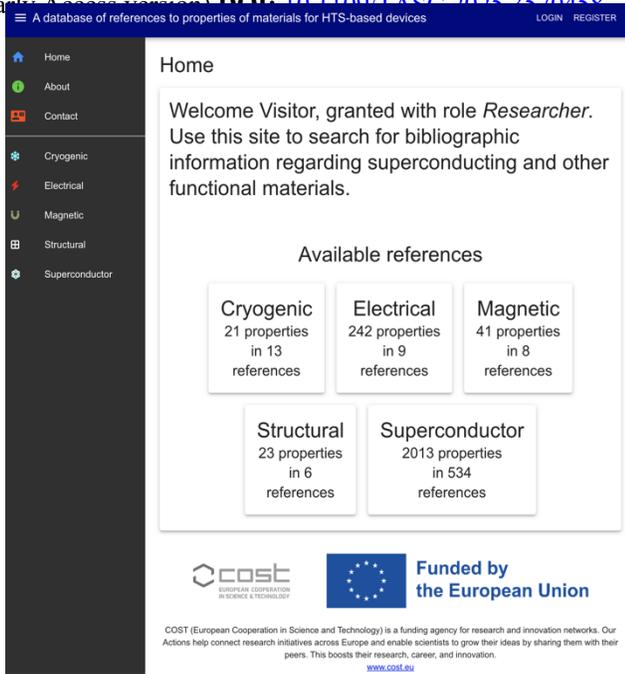

**Fig. 1.** Homepage view of the developed database available at https://sc.hi-scale.grisenergia.pt/app

community.

### III. Data Classification

The classification of data within the HTS database is designed to meet the diverse needs of researchers and engineers working on HTS technologies. By categorizing data based on material function, material type, and property characteristics, the database provides a structured approach to data retrieval, enabling users to efficiently locate specific information. This structured classification ensures that the database is a versatile resource, supporting a wide range of HTS-related applications, from mechanical support structures to advanced magnetic devices.

The classification scheme is organized into five main categories, as already mentioned: structural, cryogenic, electrical, magnetic, and superconducting materials. Each category is further divided into specific material types and targeted properties, creating a comprehensive framework that covers both HTS and auxiliary materials essential for applications.

This data classification system enables users to filter and search for data based on material type, specific material, properties, and measurement conditions. Each category is accessible through distinct tabs in the web application, with data organized into columns that correspond to publication details, material and property data, and measurement conditions, as a minimum, but with further metadata fields/columns accessible to contributors and users by a 'Selecting Fields' button, that allows to customize the size of the table to the user demands or needs through a dynamic interface. This uniform and intuitive layout ensures that users can quickly find the information they need, regardless of the material or application in question.

#### A. Structural Materials

Structural materials play a critical role in the mechanical stability and robustness of HTS devices, as they provide the support necessary for superconducting components to operate under various stress conditions. This category includes materials typically used for structural reinforcement, such as fiber-reinforced composites, metallic alloys, ceramics, and polymers. Within each type, subcategories are further defined based on specific materials, each accompanied by relevant mechanical and thermal properties.

For instance, fiber-reinforced composites in this category may include carbon fiber and glass fiber-reinforced materials, whereas metallic alloys might list brass or stainless steel. The primary properties documented for these materials are impact strength, tensile fracture strength, tensile modulus, tensile strength, and thermal conductivity. In the specific case of structural materials, all the different types of materials share the same possible properties. When additional or less common structural materials are included, entries are classified under an "Other" category to maintain classification clarity. The use of a "Other" option applies also to materials and properties.

#### B. Cryogenic Materials

This category covers materials used for thermal insulation, heat transfer, and cryogenic cooling, which are essential for maintaining the low-temperature conditions required by HTS applications. Materials in this category are divided into fluids and insulators and heat transfer materials, see Fig. 2. Cryogenic fluids such as liquid helium (LHe), liquid nitrogen (LN2), and liquid neon (LNe) are documented with properties such as density, latent heat of vaporization at boiling point, and specific heat. For insulators and heat transfer materials, properties like the coefficient of linear thermal expansion, density, dielectric strength, Poisson's ratio, thermal conductivity, and Young's modulus are included.

#### C. Electrical Materials

Electrical materials are integral to the functioning of HTS applications, as they enable or enhance the transfer of electrical currents while minimizing resistance. This category covers both conductive and dielectric materials, which serve distinct functions in electrical systems involving HTS components. The conductor subcategory includes metals like aluminum, copper, and silver, which are essential for electrical connectivity within HTS systems. Dielectric materials, such as epoxy, polyimides, and Mylar, are included for their insulating properties, preventing unintended current leakage and ensuring operational safety. Properties recorded for conductors focus on their thermal and electrical behavior, including the coefficient of linear thermal expansion, electrical conductivity, electrical resistivity, heat capacity, Poisson's ratio, specific heat, thermal conductivity, and Young's modulus. For dielectrics, the database captures properties essential to electrical insulation, such as breakdown voltage, permittivity (absolute and/or relative), and other thermal and mechanical properties.

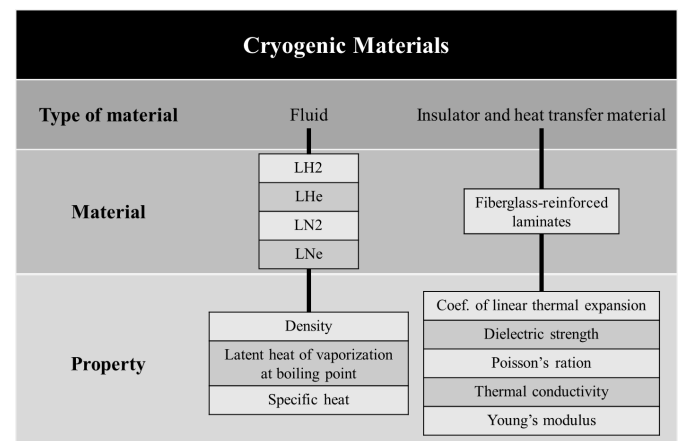

**Fig. 2.** Data model for cryogenic materials. The option "Other" is not shown as it is repeated in all levels.

*D. Magnetic Materials*

Magnetic materials are key components in HTS applications, particularly in fields like magnetic resonance imaging (MRI), magnetic levitation, and high-field magnet systems. This category focuses on materials that either generate or respond to magnetic fields, allowing for the design of complex magnetic environments needed in advanced HTS applications. They are divided into two main subcategories: permanent magnets and soft magnetic materials.

Each material in this category is characterized by properties relevant to its magnetic function and integration into HTS technologies. Permanent magnets, such as neodymium-iron-boron (NdFeB) and samarium-cobalt (SmCo), are documented with properties like the B-H curve, BH product, coercive field strength, and magnetic remanence. Soft magnetic materials, which include options like Metglas, permalloy, and silicon steel, are characterized by properties such as electrical conductivity, magnetic saturation, relative permeability, and specific losses. These properties are critical for designing HTS devices that require precise magnetic control, such as magnetic shielding or flux-pinning applications.

*E. Superconductors*

The superconductors category constitutes the core of the database and encompasses a wide range of high-temperature superconducting materials. Given the diversity in HTS materials, this category is further subdivided into several specific material types, including bulk superconductors, coated conductors, granular wires, single crystals, and thin films. Each type is documented with properties crucial for evaluating and optimizing superconducting performance.

Materials in this tab include well-known superconductors classified into custom families like BSCCO, GdBCO, YBCO, $MgB_2$, and Fe-based superconductors, among others, see Fig. 3. Each one of these families, especially in the case of composites, can in turn include numerous variants depending on fabrication and processing conditions. Then, when available from the original sources, specifics for stoichiometry, dopant type and concentration, fabrication methods (e.g., PLD, MOCVD, IBAD, RABiTS, powder-in-tube), and form of material (e.g., crystal, thin film, coated conductor, bulk), have been generally incorporated within the metadata fields/columns titled "Comments on the material" and "Additional comments".

The properties tracked for these materials are extensive and designed to meet the specific needs of HTS modelling, design, and performance optimization. They include data for measured self-field and in-field critical current density, critical temperature, higher and lower critical fields, magnetic susceptibility, pinning force or pinning force density, $n$-value, coherence length, London penetration depth, Ginzburg-Landau parameter, and anisotropy effects. Also, mechanical and thermal properties such as tensile strength, compressive strength, Young's modulus, specific heat, thermal conductivity, thermal expansion coefficient, and hardness are considered.

IV. CONCLUSION

This work presents a collaborative database of references to properties of materials for HTS-based devices available at

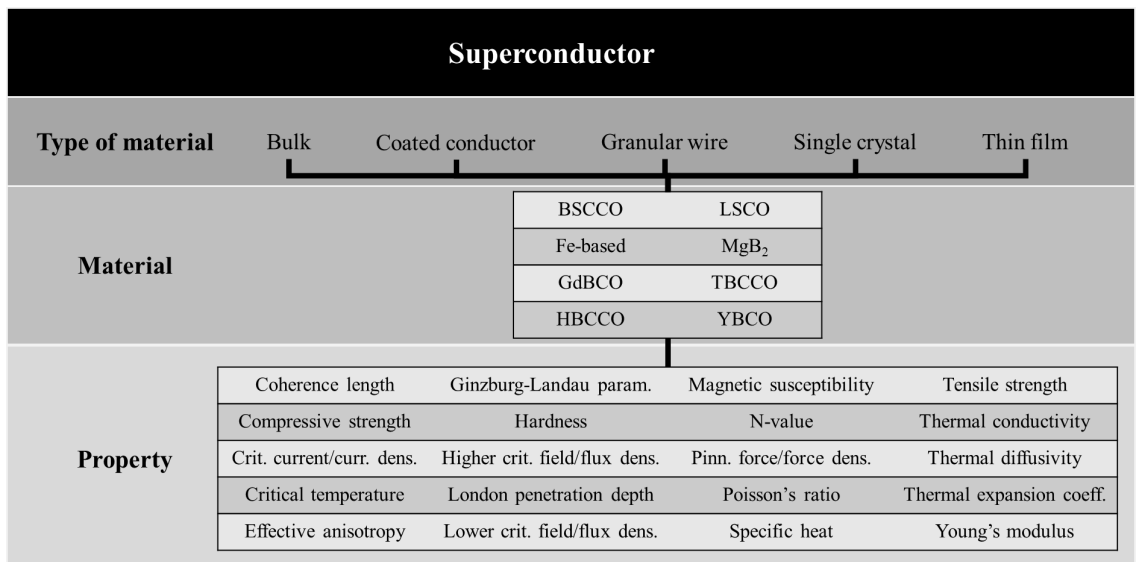

**Fig. 3.** Data model for superconductors. The option "Other" is not shown as it is repeated in all levels.

https://sc.hi-scale.grisenergia.pt/app .This database aims to address critical gaps in the accessibility and standardization of high temperature superconductors and auxiliary material data, supporting advancements in HTS technology. It provides a centralized resource for engineers, researchers, and designers in the HTS community, meeting the need for reliable, organized, and easily retrievable material properties that are essential for HTS applications.

The systematic process of gathering, verifying, and organizing the relevant data was carried out through a collaborative effort under the COST Action CA19108 initiative. Contributions were made by researchers who provided peer-reviewed data from reputable publications, ensuring a foundational level of quality and traceability. The collected data spans a wide range of material types and properties relevant to HTS applications, facilitating efficient data retrieval and comparison.

The intuitive web interface enables users to filter and search data by various parameters, including material type, properties, and measurement conditions. This user-centric design allows for easy navigation through the extensive dataset, ensuring that users can quickly access the information they need to support their research and development activities.

A distinguishing feature of this database is its community-driven nature, allowing users to actively contribute new data entries, which are then subject to peer review and verification by authorized contributors. This approach ensures continuous growth of the database while maintaining high data integrity and relevance for the HTS community. The feedback mechanisms and user support resources included in the platform further encourage engagement and ensure a smooth user experience.

By addressing the current limitations in HTS data availability and promoting a collaborative knowledge-sharing environment, this database aims to promote significant steps forward for the superconductivity community. It provides an open tool that supports both current needs and future advancements in HTS materials and applications, ultimately contributing to enable the broader adoption of HTS technologies.